\documentclass[
 reprint,
 amsmath,amssymb,
prb
]{revtex4-2}

\usepackage{graphicx}
\usepackage{dcolumn}
\usepackage{bm}

\usepackage{graphicx}
\graphicspath{ {./Figures/} }
\usepackage{xfrac}
\usepackage{amsmath}
\usepackage{amssymb}
\usepackage{amsfonts}
\usepackage{caption}
\usepackage{subcaption}
\usepackage{float}
\usepackage{color}
\usepackage{mathtools}
\usepackage{hyperref}
\usepackage{url}
\usepackage{bm}
\usepackage{eqnarray}

\newcommand*\dif{\mathop{}\!\mathrm{d}}

\usepackage{amsthm}

\usepackage[table,xcdraw]{xcolor}

\usepackage[makeroom]{cancel} 
\hypersetup{
  colorlinks   = true, 
  urlcolor     = blue, 
  linkcolor    = blue, 
  citecolor    = red 
}
\usepackage{braket}

\begin{document}




\title{Gauge-invariant optical selection rules for excitons}


\author{Tharindu Fernando}
 \email{tharindu@uw.edu}
 \affiliation{
 Department of Physics, University of Washington, Seattle, WA 98195 USA
}


\begin{abstract}
Presently, the optical selection rules for excitons 
under circularly-polarized
light are manifestly gauge-dependent.
Recently, Fernando et al. introduced a 
gauge-invariant, quantized interband index.
This index may improve the topological classification of material excited states because it is intrinsically an inter-level quantity. 
In this work, we expand on this index to introduce its 
chiral formulation, and 
use it to make the selection rules
gauge-invariant.
We anticipate this development to strengthen  
the theory of quantum materials, especially two-dimensional semiconductor photophysics. 
\end{abstract}

\maketitle







\section{Introduction}


Excitons in semiconductors are excited states
with electron-hole pairs bound by mutual Coulomb interactions
\cite{wannier1937structure,cohen2016fundamentals}.
Their ability to mediate light-matter interactions 
allows them to play significant roles in phenomena such as
optoelectronic, spintronic, and valleytronic properties 
in a wide array of material systems at the forefront of 
contemporary 
condensed matter research, particularly two-dimensional (2D)
semiconductors \cite{
koch2006semiconductor,mfc2022theoretical,mak2018light,chu20212d,latini2019cavity,selig2018dark,xiao2017excitons,yu2015valley,quintela2023anisotropic}.
Example systems include 
materials such as hexagonal boron nitride (hBN) \cite{pedersen2015intraband,cudazzo2016exciton,
galvani2016excitons,koskelo2017excitons}, 
transition-metal dichalcogenides (TMDs)
\cite{gruning2014second,trolle2015observation,gao2016dynamical,
qiu2016screening,olsen2016simple},
gapped graphene systems \cite{pedersen2009optical,cao2018unifying,
zhang2018optical,ju2017tunable},
and
2D photonic devices such as light-emitting diodes (LEDs) and lasers \cite{xiao2017excitons}.
Excitonic effects can have a significant influence on the optical
response of solids \cite{leung1997electron,albrecht1998ab,
benedict1998optical,rohlfing2000electron,onida2002electronic},
which can provide valuable information about 
material properties such as band structure \cite{boyd2008nonlinear,yu2010fundamentals},
and act as a 
sensor for probing charge ordering and quantum phases \cite{chang2024diagrammatic}
(e.g., Mott insulators in twisted bilayer moir\'e
superlattices \cite{xu2022tunable}, 
and fractional quantum Hall states in twisted bilayer 
MoTe$_2$ \cite{cai2023signatures}).
Excitons therefore offer a lucrative platform 
to study
fundamental quantum phenomena, along with applications in 
tunable materials devices
such as future excitonic-carrier devices in quantum computation and excitonic circuits \cite{xiao2017excitons}.
Therefore, it is important to have a 
unified theoretical framework to determine 
excitonic properties, which 
can offer important insights for experiments and
device applications.
A crucial step in predicting the excitonic response
is determining whether a bright or dark exciton is formed 
in response to light. 
This is determined by the optical selection rules for excitons.

The original optical selection rules for excitons
\cite{elliott1957intensity}
can be explained by the hydrogen model, since the
constituent
electron-hole pairs form
hydrogen-like bound states \cite{wannier1937structure,cohen2016fundamentals}.
For conventional semiconductors, these rules 
would state that 
in dipole-allowed materials like TMDs, 
s-like excitons are optically active, while p-like
excitons are optically inactive;
and that 
in dipole-forbidden materials, the optically active excitons are p-like states, while s-like states are optically inactive
\cite{cao2018unifying}.
However, it was recently shown that for 2D systems,
the states near the band edge may be of
multiple orbital and spin components, and the bands can
have nontrivial topological characteristics
arising from band topology (including Berry phase effects) \cite{cao2018unifying,zhang2018optical,xiao2010berry}. 
This nontrivial topology was accounted for 
by Refs. \cite{cao2018unifying,zhang2018optical}
when they proposed 
new optical selection rules for excitons under circularly
polarized light. 
These modified rules build on the fact that
the optical response of excitons 
depends on the oscillator strength
(which relies on the exciton envelope function), and the interband velocity matrix element 
characterized by angular momentum. 
Then, whether a bright or dark exciton shows for a given angular momentum 
is governed by the selection rule:
\begin{equation}\label{eq:original_selection_rules}
    m = -l_{\mp} \quad (\text{mod } n),
\end{equation}
where $n$ is from the system's discrete $n$-fold rotational
symmetry,
$m$ is the angular momentum quantum number, and
$l_\pm$ is the winding number of the interband dipole
matrix element ${\Braket{{m} | \hat{p} | n}}\cdot\hat{e}_{\pm}$ in left $(+)$ and right $(-)$
circular polarization, where $\hat{p}$ is the 
momentum operator and 
$\hat{e}_{\pm}=\frac{1}{\sqrt{2}}(\hat{e}_{k_x} \pm i \hat{e}_{k_y})$
are the chiral unit vectors, for
unit vectors 
$\hat{e}_{k_x}$ and $\hat{e}_{k_y}$ in 
momentum space $\boldsymbol{k}=(k_x,k_y)$.
These new selection rules Eq. \eqref{eq:original_selection_rules}
have since been
experimentally verified (e.g., in  Ref. \cite{ju2017tunable}),
and are key to understanding the photophysics of semiconductors.

Yet, despite their success, these modified selection rules 
may arguably be considered incomplete from a theoretical standpoint due to their inherent gauge-dependence,
which arises from requiring a hydrogenic gauge that 
doesn't allow singularities at the band edge
\cite{cao2018unifying,zhang2018optical}.
This gauge-dependence 
might impose limitations on the scope of theoretical
calculations involving the selection rules,
and therefore impact 
the potential to verify experiment and propose new physical applications.
We address this issue in this work
by making the selection rules 
gauge-invariant.
We do this using the interband index $\Theta_k$ Eq. \eqref{eq:theta:k}
for fully-gapped 2D,
Hermitian quantum systems that was proposed in Ref. \cite{fernando2023quantized,xu2018quantized}.
We introduce and use its chiral formulation $\Theta_\pm$ 
Eq. \eqref{eq:theta:chiral} to make the selection rules 
Eq. \eqref{eq:original_selection_rules} into gauge-invariant
selection rules Eq. \eqref{eq:excitons:new:rule}. 
This gauge-invariance 
may strengthen the theory of excitonic optical selection rules,
thereby making associated phenomenology more accessible for 
both theoretical and experimental work.
\\


\section{Quantized, gauge-invariant interband index}
\begin{figure*}
    \centering
    \includegraphics[width=0.9\textwidth]{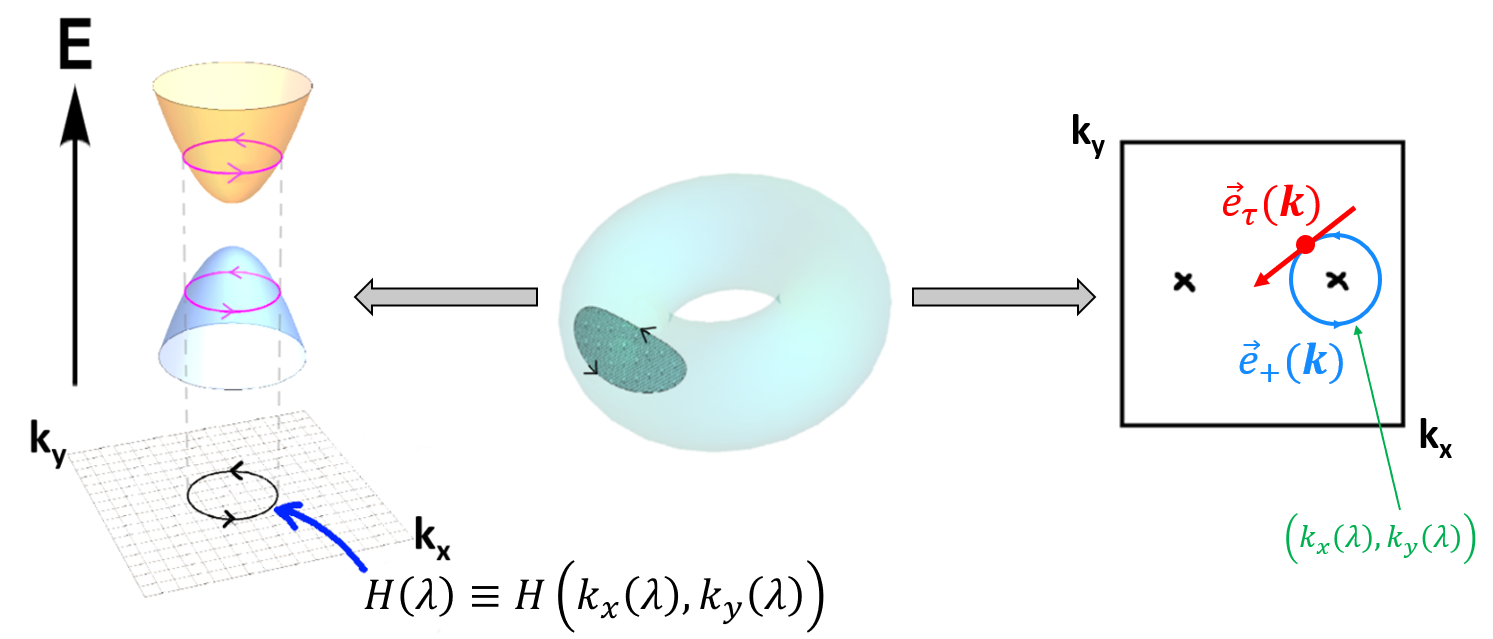}
    \caption{    
        \emph{Center:} Two-dimensional $k$-space Brillouin torus 
        for a two-level system. 
        The counterclockwise loop is $\partial\mathcal{M}$,
        and defines
        the shaded region of the torus
        as $\mathcal{M}$, by convention. 
        \emph{Left:} Two bands of the dispersion $E$
        (valence and conduction bands of a material system, for example)
        in the vicinity of a band edge.
        If the $k$-space loop is parameterized by
        $\lambda$, the expressions ${H}(\lambda)\equiv  {H}(k_x(\lambda), k_y(\lambda))$ are equivalent descriptions 
        of the adiabatic loop $\partial\mathcal{M}$ 
        shown on the Brillouin zone under $E$.
        \emph{Right:} For $\boldsymbol{k}$ constrained to a closed loop 
        $\partial\mathcal{M}$,
        the tangential vector $\vec{e}_\tau (\boldsymbol{k})$
        at a point
        is denoted in red,
        while the counterclockwise chiral unit vector $\vec{e}_{+} (\boldsymbol{k})$ for $\Theta_{+}$ may be 
        visualized by the same $\partial\mathcal{M}$
        (with the opposite orientation for $\vec{e}_{-} (\boldsymbol{k})$).
        The $\textbf{X}$'s indicate example 
        band edge points.
    }
    \label{fig:intro:interband}
\end{figure*}

Drawing from Ref. \cite{fernando2023quantized}, consider the time-independent Schr\"{o}dinger equation for 
an $N$-level non-degenerate Hamiltonian $H(\boldsymbol{k})$:
\begin{equation}
\label{eq:Theta:instantaneous}
    {H}(\boldsymbol{k})\ket{m(\boldsymbol{k})} = E_m (\boldsymbol{k}) \ket{m(\boldsymbol{k})},
    (m = 1,2, ... , N),
\end{equation}
where $\ket{m(\boldsymbol{k})}$ are orthonormal instantaneous eigenstates associated with eigenvalues $E_m (\boldsymbol{k})$.
Then, we define the interband index $\Theta_k$: 
\begin{align}
\begin{split}
	\label{eq:theta:k}
	2\pi\Theta_{\boldsymbol{k}} &= \Delta\Phi
	- \oint\limits_{\textrm{ }\partial{ \mathcal{M}}}
	\dif \arg\Braket{{m} | \boldsymbol{\nabla_k} n}\cdot
	\hat{e}_{\tau} \\
	&=\Delta\Phi
	- \oint\limits_{\textrm{ }\partial{ \mathcal{M}}}
	\dif \arg\frac{\Braket{{m} | \boldsymbol{\nabla_k} H | n}}{E_{nm}}\cdot
	\hat{e}_{\tau}, 
	\end{split}
\end{align}
where we used the Hellman-Feynman-type relation $\Braket{{m} | \dif_\lambda n}
= \Braket{{m} | {\dif_\lambda H}/{E_{nm}} |n}$ (for $\ket{m}\neq\ket{n}$)
in the last equality.
Above, we used the definitions: 
 $\dif$ is the total derivative with respect to $k_x$ and $k_y$;
$\boldsymbol{\nabla_k}=(\partial_{k_x},\partial_{k_y})$;
$\hat{e}_{\tau}={\boldsymbol{\dot{k}}}/{|\boldsymbol{\dot{k}|}}$ 
is the unit tangential vector at a point
on the loop $\partial{ \mathcal{M}}$ (see Fig. \ref{fig:intro:interband}); 
$\boldsymbol{\dot{k}}={\dif{\boldsymbol{k}(\lambda)}}/{\dif{\lambda}}$ for some $\lambda$ that
parameterizes the loop $\boldsymbol{k}=(k_x(\lambda),k_y(\lambda))$;
and $\Delta\Phi_{mn}=\Phi_{m} - \Phi_{n}$,
where $\Phi_m=\int_{\partial{\mathcal{M}}}
	 \mathcal{A}_{m}^\mu \dif\lambda_{\mu}
	 -
	 \iint_{{\mathcal{M}}}
	 F_m\dif \lambda_{\mu}\dif\lambda_{\nu}.$
For brevity, we henceforth drop the 
differential elements $\dif\lambda_{\mu}$.
$\Phi_m$ is the number of
\emph{Berry singularities} in level $|m\rangle$:
It
is the difference between 
the line integral of the standard 
Berry connection
$\mathcal{A}_{m}^\mu = i \braket{{m} | \frac{\partial}{\partial \lambda_{\mu}} m}$
along $\partial{ \mathcal{M}}$,
and the area integral of the 
Berry curvature $F_m = \frac{\partial}{\partial \lambda_{\mu}} \mathcal{A}_{m}^\nu 
	 - 
\frac{\partial}{\partial \lambda_{\nu}} \mathcal{A}_{m}^\mu$
over the region $\mathcal{M}$ specified by the loop.
Therefore, $\Phi_m$ is the
quantized `amount' by which
Stokes' theorem fails.
Then, $\Delta\Phi$ is the net number of Berry singularities
between the levels considered. 
Notice
that in the case without gauge singularities,
$\Phi_m $ reduces to $0$ as $\int_{\partial{\mathcal{M}}}\mathcal{A}_{m}^\mu =\iint_{{\mathcal{M}}}F_m$.

The $k$-dependent form in Eq. \eqref{eq:theta:k}
makes it clear that the vector $\hat{e}_{\tau}$ 
may be replaced by an appropriate vector that preserves a
winding property of the matrix element $\Braket{{m} | \boldsymbol{\nabla_k} H | n}$.
Therefore, we introduce another physically-useful formulation of the interband index 
in terms of the chiral unit vectors $\hat{e}_{\pm}$ (see Fig. \ref{fig:intro:interband}):
\begin{equation}
	\label{eq:theta:chiral}
	2\pi\Theta_{\pm} = \Delta\Phi
	- \oint\limits_{\textrm{ }\partial{ \mathcal{M}}}
	\dif \arg\frac{\Braket{{m} | \boldsymbol{\nabla_k} H | n}}{E_{nm}}\cdot
	\hat{e}_{\pm}.
\end{equation}
All the terms in the definitions of the interband index work together to 
give a quantized, gauge-invariant quantity.
\\

\section{Gauge-invariant optical selection rules for excitons}
The chiral unit vectors 
$\hat{e}_{\pm}$
are known to decompose
the interband matrix element ${\Braket{{m} | \hat{p} | n}}$ into chiral components 
${\Braket{{m} | \hat{p} | n}}\cdot\hat{e}_{+}$ and 
${\Braket{{m} | \hat{p} | n}}\cdot\hat{e}_{-}$.
These may respectively correspond to left- and right-circularly polarized
photon modes ($\sigma_{-}$ and $\sigma_{+}$) \cite{cao2018unifying}. 
This means that the complex-valued 
${\Braket{{m} | \hat{p} | n}}\cdot\hat{e}_{\pm}$ yield two unique vector fields 
with possibly different winding patterns over $k$-space.
Since $\hat{p}$ is the $k$-space derivative of the Hamiltonian,
we observe that the winding number of ${\Braket{{m} | \hat{p} | n}}\cdot\hat{e}_{\pm}$ 
is simply
the interband line integral in our chiral formulation of the interband index Eq. \eqref{eq:theta:chiral}.
Therefore, we rid the optical selection rules Eq. \eqref{eq:original_selection_rules}
of their explicit gauge-dependence by 
rewriting them as:
\begin{equation}
	\label{eq:excitons:new:rule}
	m = -\Theta_{\mp} \quad (\text{mod } n).
\end{equation}
One could argue that the above selection rule 
Eq. 
\eqref{eq:excitons:new:rule} strengthens
the theory of excitons 
due to explicit gauge-invariance.

We computationally
verified Eq. \eqref{eq:excitons:new:rule}  by calculating $\Theta_\pm$ (and $m$) for the biased bilayer graphene
model that is used as an example in Ref. \cite{zhang2018optical}.  
From Ref. \cite{zhang2018optical,park2010tunable,mccann2006landau}:
\begin{equation}\label{eq:BLG:ham}
H = \begin{pmatrix}
\Delta & \alpha k_{+}^{2} \\
\alpha k_{-}^{2} & -\Delta
\end{pmatrix}
+ 3\gamma_{3} \begin{pmatrix}
0 & k_{-} \\
k_{+} & 0
\end{pmatrix},
\end{equation}
where $k_{\pm}=k_x \pm i k_y$, $\gamma_3$ is the interlayer
hopping amplitude, $2\Delta$ is the energy gap, and we set 
the factor $\alpha = 1$.

To calculate the winding number given by the 
line integral in Eq. \eqref{eq:theta:chiral},
we numerically discretized a circular counterclockwise loop 
(see Fig. \ref{fig:intro:interband}) around the band edge point $\boldsymbol{k}=(0,0)$, and used 
finite differences (a central difference) to compute the derivatives of the Hamiltonian $H(\boldsymbol{k})$.
To calculate $\Phi_m$, we discretized the
two-dimensional $k$-space grid in the region inside the 
circular loop, and computed for each energy level $\ket{m}$ the 
phases from both the Berry curvature $F_m$ and Berry connection
$\mathcal{A}_{m}$ integrals separately. 
For the $F_m$ integral, 
we used the standard numerical technique 
given in Ref. \cite{fukui2005chern} to determine the phase of the product of overlap matrices around each closed rectangular path
of the $k$-space grid, but only for the rectangles 
inside the circular loop. 
The $\mathcal{A}_{m}$ integral was obtained by summing the phases of the individual overlap matrices along each segment of each closed rectangular path. 
We used \emph{Python} and \emph{MATLAB}
for our numerical simulations. 
Separately, we verified our results
using analytic expressions for wavefunctions and derivatives
using the Hamiltonian Eq. \eqref{eq:BLG:ham}, using 
\emph{Mathematica}.

Without the interlayer hopping term (i.e. $\gamma_3 = 0$), the 
biased bilayer graphene system has $C_\infty$ symmetry
and yielded $\Theta_{+} = -3$ and $\Theta_{-} = -1$ (i.e. $m=3$ and $m=1$ from Eq. \eqref{eq:excitons:new:rule}), 
as expected in Ref. \cite{zhang2018optical}.
Non-zero $\gamma_3$ reduces $C_\infty$ to $C_3$ and should give $m=-2,0,1,3=0,1\text{ (mod 3)}$.
Using $\gamma_3=0.3$,
we got $\Theta_{+} = 0$ and $\Theta_{-} = -1$ (i.e. $m=0$ and $m=1$), as expected.
However, we caution against using this reasoning 
to determine which excitons show,
without first calculating
the oscillator strength explicitly. This is because the mapping involving $\Theta_\pm$ may be many-to-one, 
while the full calculation might specify a many-to-many relation in the sense that different
$m$ may result from the same $\Theta_\pm$ (as we saw 
when mapping $m=-2,0,1,3\rightarrow 0,1$).
\\

\section{Conclusion}
We showed how the interband index
$\Theta_\pm$ Eq. \eqref{eq:theta:chiral} may be used to make 
the gauge-dependent optical selection rules for excitons
Eq. \eqref{eq:original_selection_rules}
gauge-invariant Eq. \eqref{eq:excitons:new:rule}.
This gauge-invariance may broaden the scope
of the types of theoretical calculations that may be
done, since we now 
do not need to fix the gauge or transform between 
different gauges.
This may lead to fewer ambiguities in physical interpretations that stem from gauge-dependence, since 
we now have a unified theoretical framework
that is consistent across different materials and models.
By this enhancement of the predictive power of theoretical models,
our new selection rules may now
allow for clearer physical insights into excitonic effects,
since they may allow for more-accurate comparisons
between theoretical simulations and experiments 
(which are typically gauge-invariant). 
This may lead to potentially new phenomena
and advances in material design (especially of quantum 
materials possessing novel electronic and optical properties),
which may be achievable in the near-term 
given today's high rate of advances in 
interband physics and topological
effects in systems like
multilayer graphene, TMDs,
Moir\'e heterostructures, qubits,
and even other kinds of quasiparticles that can couple with each other (e.g., excitons, magnons, polaritons, polarons, plasmons, and phonons).
These advantages may directly influence the development
of advanced materials and technologies, with potential applications in
optoelectronics, photovoltaics, quantum computing, excitonics, spintronics, and other exciting research fronts in contemporary materials science and condensed matter physics.
\\

We acknowledge helpful discussions with Ting Cao and Di Xiao 
at the University of Washington. 







\end{document}